\documentclass[twocolumn,showpacs,preprintnumbers,amsmath,amssymb]{revtex4}
\usepackage{tabularx,graphicx}\begin{document}
\newcommand{\beq}{\begin{equation}}
\newcommand{\eeq}{\end{equation}}
\newcommand{\beqn}{\begin{eqnarray}}
\newcommand{\eeqn}{\end{eqnarray}}
\newcommand{\bmath}{\begin{subequations}}
\newcommand{\emath}{\end{subequations}}
\title{Charge expulsion and electric field in superconductors}
\author{J. E. Hirsch }
\address{Department of Physics, University of California, San Diego\\
La Jolla, CA 92093-0319}
 
\date{August 31, 2003} 
\begin{abstract} 
The theory of hole superconductivity predicts that when a metal goes
superconducting negative charge is expelled from its interior towards the surface. 
As a consequence the superconductor in its ground state is predicted to have a 
non-homogeneous charge distribution 
and an outward pointing electric field in its interior. 
Here we propose equations to describe the behavior of the charge density and electric field 
in superconductors, and solve them for a spherical geometry.  The magnitude of the predicted interior electric field 
depends on superconducting parameters such as the condensation energy and
the London penetration depth and is found to be of order $10^6 V/cm$.
A physical interpretation of the result is given. 
It is predicted that for small superconducting bodies (compared to the
penetration depth) an electric field $outside$ the superconductor should result from this
physics. This may explain a recent experimental observation in $Nb$ metal clusters.
\end{abstract}
\maketitle

\section{Introduction}

In the currently accepted understanding of superconductivity no electric fields exist in  
superconductors in the absence of electric current (a 'Bernoulli potential' is expected
to exist in the presence of a non-uniform supercurrent\cite{bernoulli}). The 
London brothers\cite{london1} originally proposed a set of
equations to describe superconductivity that allowed for the presence of an electric field
within a penetration depth of the surface of a superconductor. However H. London
failed to detect such an electric field experimentally\cite{london2}. Having confidence in
his brother's experimental result, F. London modified the theory and discarded one
of their original equations, and as a consequence the 
possibility of an electric field in superconductors is no longer discussed in
London's  definitive work\cite{london3}.

As is well known, normal metals allow for the existence of magnetic fields but no
electric fields in their interior. Superconductors (type I) do not allow magnetic
fields in their interior, but magnetic fields can exist within a penetration depth of the
surface. Based on the theory of hole superconductivity\cite{hole,hole2} we have 
recently proposed that when
a metal goes superconducting negative charge is expelled from its interior towards
the surface\cite{charge}. Here we discuss the equations governing the
charge and electric field distribution in superconductors resulting from this
physics. The possibility that a superconductor may have an electric field in its 
interior has not been discussed in other theoretical frameworks to our knowledge; 
it is however a necessary consequence of the fundamental electron-hole asymmetry
of condensed matter\cite{ehasym}, and the resulting 'giant atom' description of 
superconductors\cite{atom} 
that results from the theory of hole superconductivity.

\section{Energetics}

\begin{figure}
  \includegraphics[height=.25\textheight]{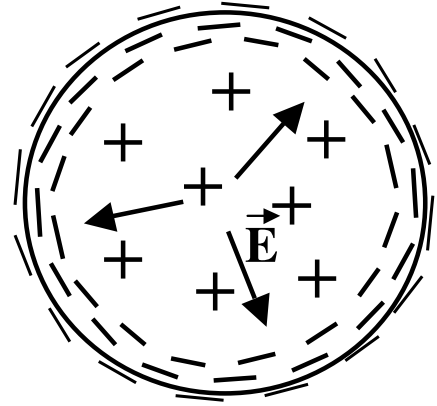}
  \caption{Schematic picture of a spherical superconducting body. Negative
charge is expelled from the bulk to the surface. As a consequence, an outward-pointing
electric field exists
in the interior.}
\end{figure}

The qualitative picture of a superconductor proposed in refs. \cite{charge,atom} is
shown in figure 1: it looks like a giant atom, with a higher density of negative
charge near the surface and higher density of positive charge in the
interior\cite{explain}. Consider a superconducting sphere of radius $R$ much larger than its
London penetration depth $\lambda_L$. Assume an amount of charge $q$
moves from the interior to the surface, resulting in a uniform charge
distribution $\rho_0$ in the interior:
\beq
\rho_0=-\frac{3q}{4\pi R^3}
\eeq
As we discuss later, $q$ is negative and resides within a London penetration
depth of the surface, with volume charge density
\beq
\rho_{-}=\frac{q}{4\pi R^2\lambda_L}=-\frac{1}{3} \frac{R}{\lambda_L}\rho_{0}
\eeq
For $\lambda_L<<R$ we can also think of this charge as a 'surface charge
density'
\beq
\sigma=\frac{q}{4\pi R^2}=\rho_{-}\lambda_L
\eeq
This 'surface charge density' is $not$ confined within a Thomas Fermi screening length
of the surface as excess charge would be in ordinary metals, but within the much thicker surface layer
defined by the penetration depth.

The electric field in the interior is given by
\beq
\vec{E}(r)=-\frac{q}{R^3}\vec{r}
\eeq
and the inhomogeneous charge distribution gives rise to a cost in Coulomb energy
\beq
U_E=\int_0^R d^3r \frac{E(r)^2}{8\pi}=\frac{q^2}{10R}
\eeq
under the assumption $\lambda_L<<R$. For a spherical geometry no electric field exists outside the
superconductor (assuming charge neutrality).

In the superconducting state, the energy per unit volume is lowered by the superconducting
condensation energy. The theory of hole superconductivity predicts that the 
charge expulsion describe above is a necessary consequence of superconductivity.
Hence the Coulomb energy cost Eq. (5) is balanced by an energy gain due to the
charge expulsion,  $\tilde{\epsilon}$ per unit volume,
which is related but not necessarily equal to the condensation energy of the superconductor.
Equating this energy gain to the electrostatic energy cost,
\beq
\frac{q^2}{10R}=\frac{4}{3}\pi R^3 \tilde{\epsilon}
\eeq
yields for the 'surface charge density'
\beq
\sigma=[\frac{5}{6 \pi} \tilde{\epsilon}]^ {1/2}
\eeq
independent of the sphere's radius. This indicates that if the
physics discussed here indeed exists in superconductors, the
charge density in the surface layer $\rho_{-}$ is an intensive quantity determined by
intrinsic properties of the superconductor, in particular the penetration depth
and the thermodynamic critical field $H_c$ (which gives the condensation energy).
Instead, the interior positive charge density $\rho_0$ is not intensive but
decreases with increasing radius $R$. The electric field close to the
surface, given by Eq. (4) for $r\sim R$, is also independent of the
body's dimensions.

From Eqs. (3) and (7) the volume charge density in the surface layer is given by
\beq
\rho_-=-[\frac{5}{6\pi \lambda_L^2}\tilde{\epsilon}]^{1/2}
\eeq

Using the expression for the London penetration depth
\beq
\frac{1}{\lambda_L^2}=\frac{4\pi n_s e^2}{m_ec^2}\eeq
with $m_e$ the free electron mass, $e$ the electron charge (negative) and $n_s$ the superfluid
density per unit volume, and replacing $\tilde{\epsilon}$ in terms of 
the energy gained per superfluid electron $\epsilon$ yields
\beq
\epsilon=\frac{\tilde{\epsilon}}{n_s}
\eeq
and Eq. (8) becomes
\beq
\rho_-=en_s(\frac{10}{3}\frac{\epsilon}{m_ec^2})^{1/2}
\eeq
relating the negative excess charge density in the surface layer to the superfluid density $n_s$ and the
ratio of the condensation energy of an electron $\epsilon$ to its rest energy,
showing that the excess charge density is a small fraction of the
superfluid density. The 'surface charge density' $\sigma$ is given by 
Eq. (3), and the maximum electric field attained within a
penetration length of the surface of the superconductor is
\beq
E_{max}=-4\pi \sigma=-4 \pi \lambda_L \rho_{-}
\eeq
and is also independent of the sample's dimensions.

As an example appropriate for high $T_c$ cuprates we take
$\lambda_L=2000 \AA$, yielding from Eq. (9) a superfluid density
$n_s=7.1\times 10^{-4}$ electrons /$\AA^3$.
Assuming a condensation energy of $100\mu eV$ per unit cell and a 
unit cell volume of $40 \AA^3$ yields $\epsilon=3.5 meV$ per superfluid
electron,
and $\rho_-=-1.5\times 10^{-4} n_s e$ so there is roughly $1$ extra
electron per $10,000$ superfluid electrons in the outer layer. The 
'surface charge density' is $\sigma=-3.5\times 10^{-7} C/cm^2$, and the maximum
electric field near the surface is $E_{max}=3.9\times 10^6 V/cm$. The potential
difference between the center of the sphere and a point within a
penetration depth of the surface is $\Delta V=E_{max}R/2$, so
approximately $4$ million Volt for a sample of $1cm$ radius.

As another example, for Nb the thermodynamic critical field is $H_c=1980G$,
so the condensation energy is $\tilde{\epsilon}=1.56\times 10^5 ergs/cm^3$.
From Eq. (7) the 'surface charge density' is $\sigma=6.8\times 10^{-8} C/cm^2$ and 
taking $\lambda_L=400\AA$ for the penetration depth yields $\rho_-=0.017 C/cm^3$ for
the surface layer volume charge density, corresponding
to approximately $2$ excess electrons per million $Nb$ atoms in the
outer layer. The maximum electric field near the surface from Eq. (12)
is $E_{max}=0.77\times 10^6 V/cm$. This is not very different from the estimate for 
the high $T_c$ case
 because the effects of a smaller condensation energy and
a smaller London penetration depth partially compensate each other.

\section{Electric field equations}

We start with the London equation for the supercurrent
\beq
\vec{J}=-\frac{n_s e^2}{m_e c}\vec{A}
\eeq
with $\vec{A}$ the magnetic vector potential. Following the Londons\cite{london1} we
assume that $\vec{A}$ satisfies the Lorenz gauge condition
\beq
\vec{\nabla} \cdot \vec{A} + \frac{1}{c}\frac{\partial \phi}{\partial t}=0
\eeq
with $\phi$ the electric potential. The electric field is given by
\beq
\vec{E}=-\vec{\nabla}\phi -\frac{1}{c} \frac{\partial\vec{A}}{\partial t}
\eeq
Using the continuity equation
\beq
\vec{\nabla}\cdot\vec{J}=-\frac{\partial \rho}{\partial t}
\eeq
and applying the divergence operator to both sides of Eq. (13) and using the
gauge condition Eq. (14) yields
\beq
\frac{\partial \phi}{\partial t}=-\frac{m_e c^2}{n_se^2} \frac{\partial \rho}{\partial t}
\eeq
hence 
\beq
\phi (\vec{r},t)=-\frac{m_e c^2}{n_s e^2} \rho (\vec{r},t)+\phi _0(\vec{r})
\eeq
The London brothers postulated Eq. (18) with $\phi_0(\vec{r})=0$ as a possible
equation applicable to superconductors\cite{london1}. Under stationary conditions,
\beq
\phi(\vec{r})=-4\pi \lambda_L^2 \rho(\vec{r})
\eeq
and using Maxwell's equation
\beq
\vec{\nabla}\cdot\vec{E}(\vec{r})=4\pi \rho(\vec{r})
\eeq
Eqs. (15) and (19) yield
\bmath
\beq
\rho(\vec{r})=\lambda_L^2 \nabla^2\rho(\vec{r})
\eeq
\beq
\vec{E}(\vec{r})=\lambda_L^2 \nabla^2\vec{E}(\vec{r})
\eeq
\emath
which imply that a non-zero charge density and an electric field can exist within a 
penetration depth of the surface of the superconductor but not in the interior.
H. London attempted to measure the electric field predicted by Eq. (21) by measuring
the change in capacitance of a capacitor with superconducting plates when
it is cooled below $T_c$\cite{london2}. However he failed to detect any effect.

Here we propose that Equation (18) is valid for superconductors
{\it with a non-zero} $\phi_0(r)$ resulting from a uniform charge density $\rho_0$ deep
in the interior of the superconductor, as discussed in the previous section. Eqs. (21) and (18) then
become
\bmath
\beq
\rho(r)=\rho_0+\lambda_L^2 \nabla^2\rho(r)
\eeq
\beq
\vec{E}(r)=\vec{E}_0(r)+\lambda_L^2 \nabla^2\vec{E}(r)
\eeq
\beq
\phi (\vec{r})=-4\pi \lambda_L^2 \rho (\vec{r})+\phi _0(\vec{r})
\eeq
\emath
with $\vec{E}_0(\vec{r})=-\vec{\nabla}\phi_0(\vec{r})$ and 
$\vec{\nabla}\cdot\vec{E}_0(\vec{r})=4\pi \rho_0$. 
We propose that these equations   hold for the interior of
superconductors of arbitrary shape in the absence of magnetic fields and electric currents, with $\rho_0$
a positive number determined by the microscopic parameters of the superconductor
as well as the geometry, as discussed in the previous section.
The potential   obeys
\bmath
\beq
\phi (\vec{r})=\phi _0(\vec{r})+\lambda_L^2 \nabla^2\phi (\vec{r})
\eeq
\beq
\nabla^2\phi_0 (\vec{r})=-4\pi\rho_0
\eeq
in the interior, and
\beq
\nabla^2\phi(\vec{r})=0
\eeq
\beq
\nabla^2\phi_0(\vec{r})=0
\eeq
in the exterior of the body. Furthermore we assume no surface charges can exist in the superconductor,
hence that both $\phi$ and its normal derivative $\partial \phi/ \partial n$ are continuous on the
surface of the body, as are of course $\phi_0$ and  $\partial \phi_0/ \partial n$ .
\emath

Note that the existence of an electric field in the interior of the superconductor does
neither imply a time variation of the current, as one would expect in a 'perfect conductor', nor
even the $existence$ of a current as in an ordinary metal. Taking the time derivative of Eq. (13)
and using the Maxwell equation (15) yields
\beq
\frac{\partial \vec{J}}{\partial t}=\frac{n_se^2}{m}(\vec{\nabla} \phi +\vec{E})
\eeq
so that in a stationary situation there can be an electric field that is derivable
from a potential ($\vec{E}=-\vec{\nabla} \phi$) and it does not lead to a time-varying 
supercurrent. Whether a stationary supercurrent exists or not depends on the
magnetic vector potential $\vec{A}$ through Eq. (13) and not on the electric field.

In a spherical geometry, the solution of Eq. (22a) is
\beq
\rho(r)=\rho_0-k\frac{\sinh (r/\lambda_L)}{r}
\eeq
and for charge neutrality the constant $k$ is given by
\bmath
\beq
k=\frac{1}{3}\frac{\rho_0R^3}{\lambda_L^2}\frac{1}{f(R/\lambda_L)}
\eeq
\beq
f(x)=x\cosh x - \sinh x
\eeq
\emath
For $R>>\lambda_L$ the charge density at the surface is
\beq
\rho(R)=\rho_0(1-\frac{1}{3}\frac{R}{\lambda_L})\sim -\frac{R}{3\lambda_L}\rho_0
\eeq
\newline
in accordance with Eqs. (1), (2), and for $r<<R-\lambda_L$, $\rho(r)\sim \rho_0$.
The electric field is given by
\beq
\vec{E}(r)=\frac{4}{3}\pi\rho_0[1-\frac{R^3}{r^3}\frac{f(r/\lambda_L)}{f(R/\lambda_L)}] \vec{r}
\eeq
It goes to zero at $r=R$ and is maximum at a distance of a penetration depth inside the surface,
given by
\beq
E_{max}\sim \frac{4}{3} \pi \rho_0 R = -4 \pi \lambda_L \rho(R)
\eeq
in agreement with Eq. (12) for $\rho(R)=\rho_-$, as expected. 
 
\begin{figure}
 \includegraphics[height=.39\textheight]{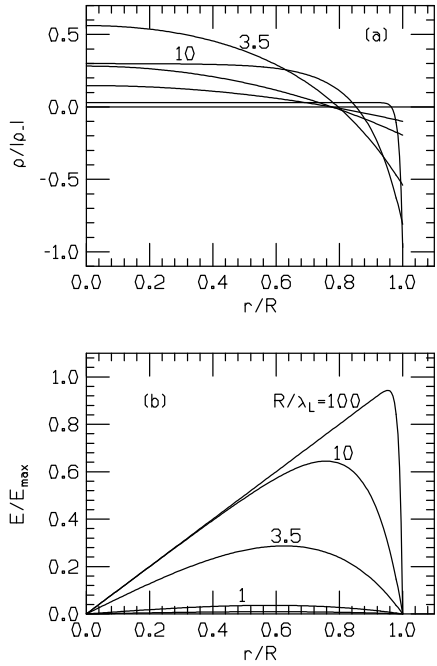}
  \caption{Radial dependence of charge density (a) and electric field (b) for  
$R/\lambda_L=0.5,1,3.5,10$ and $100$ (some numbers next to the lines).
The electric field increases monotonically with increasing $R/\lambda_L$. 
The charge density at the origin is smallest for $R/\lambda_L=100$, and decreases
with decreasing $R/\lambda_L$ for $R/\lambda_L<3.5$.}
\end{figure}

Figure 2 shows the
radial dependence of the charge density  and electric field for various values of 
$R/\lambda_L$. When the dimensions of the sample are much larger than the 
penetration depth there is a small uniform positive charge density in the interior and the
electric field increases linearly towards the surface. The compensating negative
charge resides within a penetration length of the surface and the electric field
drops to zero in that region. As the dimensions of the sample decrease the
positive charge density in the interior increases and is maximum for
$R/\lambda_L\sim3.5$, where $\rho_0\sim-0.56\rho_-$, and decreases
again for even smaller samples. This is under the assumption that 
superconducting properties are not affected by the size of the sample,
i.e. $\rho_-$ and $\lambda_L$ are assumed independent of $R$.
In the limit $R/\lambda_L<<1$ the charge density is given by
\beq
\rho(r)=\frac{\rho_0}{6\lambda_L^2}[\frac{3}{5}R^2-r^2]
\eeq
so that it changes sign at $r/R=0.77$, and the electric field by
\beq
\vec{E}(\vec{r})=\frac{2\pi}{15}\frac{\rho_0}{\lambda_L^2}[R^2-r^2]\vec{r}   .
\eeq

\section{Microscopic interpretation}

The existence of an electric field in the interior of a superconductor may seem surprising.
In normal metals, electric fields cannot exist in the interior because the mobile electronic
charge will rearrange to nullify such field. How can the existence of an electric
field in a superconductor be understood?

Consider Eq. (24). It states that no electric current will be generated in the presence
of an interior electric field, provided the field derives from a potential
$\phi$. Now an electric field exerts a force on electrons, why is it that electrons
in the superconductor do not get accelerated by the interior electric field and
generate an electric current?

The Lorentz force equation in the absence of a magnetic field
\beq
m_e\frac{d\vec{v}}{dt}=e\vec{E}
\eeq
indicates that the velocity of an electron will change in the presence of an electric
field. From the relation between velocity and current
\beq
\vec{J}=en_s\vec{v}
\eeq
one may be tempted to conclude that Eqs. (32) and (33) are incompatible
with Eq. (24). However this is $not$ the case, because Eq. (24) is a $local$ time
derivative and Eq. (32) is a $total$ time derivative.
Using the relation\cite{sommer}
\beq
\frac{d\vec{v}}{dt}=\frac{\partial\vec{v}}{\partial t}+\vec{\nabla}\frac{v^2}{2}-
\vec{v}\times(\vec{\nabla}\times\vec{v})
\eeq
we conclude that a stationary situation with an electric field in the interior of the superconductor is indeed possible,
provided that
\beq
\vec{E}=\frac{m_e}{e}[\vec{\nabla}\frac{v^2}{2}-\vec{v}\times(\vec{\nabla}\times\vec{v})]
\eeq

If we assume that the superfluid electrons traverse circular orbits with speed $v(r)$ we have
\beq
\vec{\nabla}\frac{v^2}{2}-\vec{v}\times(\vec{\nabla}\times\vec{v})=-\frac{v^2(r)}{r}\hat{r}
\eeq
and the speed is obtained from Eqs. (35) and (28) as
\beq
v^2(r)=-\frac{4}{3}\frac{\pi e \rho_0}{m_e}r^2[1-\frac{R^3}{r^3}\frac{f(r/\lambda_L)}{f(R/\lambda_L)}]
\eeq
which of course simply describes the balance between centripetal acceleration and electric Lorentz force
\beq
\frac{m_ev^2}{r}=-eE
\eeq
as appropriate for a circular orbit. Deep in the interior of the superconductor Eq. (37) describes a 'rigid rotation'
of the superfluid, $v(r)=\omega r$,   with constant angular velocity 
\beq
\omega=(-\frac{4\pi \rho_0e}{3m_e})^{1/2}
\eeq
As we approach the surface the angular velocity   decreases to zero due to the screening of the electric field by the negative
charge in the surface layer.

We conclude from this interpretation
that the 'rigidity' of the wave function\cite{london3} that allows for an electric field
in the interior arises from the fact that superfluid electrons traverse macroscopic
orbits at high speed; a radial electric field will not change the radial speed of electrons
for the same reason that an orbiting planet is not attracted towards the sun: the radial
force provides the centripetal acceleration for the orbital motion. 

However one may also ask: if there is a finite velocity as given by Eq. (37), shouldn't
there be a finite current and an associated magnetic field? In fact Eq. (33) is an
oversimplification, which assumes that all electrons at position $\vec{r}$ move with
the same velocity $\vec{v}$. However electrons also have a spin quantum number.
As the simplest resolution to this question we may assume that the velocity field for
spin direction $\sigma$ (with respect to a chosen quantization axis) obeys
\beq
\vec{v}_\sigma(\vec{r})=-\vec{v}_{-\sigma}(\vec{r})
\eeq
with $v_\sigma$ given by Eq. (37). In that case the total $charge$ current
\beq
\vec{J}=e\frac{n_s}{2}(\vec{v}_\uparrow+\vec{v}_\downarrow)
\eeq
will be zero and no magnetic field will be generated, however there will be a $spin$
current
\beq
\vec{J}_{spin}=e\frac{n_s}{2}(\vec{v}_\uparrow-\vec{v}_\downarrow)
\eeq
which is not zero. The direction of the spin current would be determined by
spin-orbit coupling as discussed in Ref. \cite{atom}.

A different simple solution of Eq. (35) is obtained under the assumption $\vec{\nabla}\times \vec{v}=0$.
In that case, we have simply
\beq
\vec{E}=\frac{m_e}{2e}\vec{\nabla}v^2
\eeq
and the velocity profile is readily obtained by integration of Eq. (28) as
\bmath
\beq
v^2(r)=\frac{4}{3}\frac{\pi e \rho_0}{m_e} r^2[1-2\frac{R^3}{r^3}\frac{\sinh (r/\lambda_L)}{f(R/\lambda_L)}]
+k 
\eeq
with $k$ an integration constant. Deep in the interior of the superconductor we have (for $R/\lambda_L >>1$) 
\beq
v^2=\frac{4\pi\rho_0 e}{3m_e} r^2+k
\eeq
\emath
Eq. (44b) has a simple physical interpretation. The potential
in which the electron moves in the interior of the superconductor is a harmonic
oscillator potential:
\beq
U(r)=e\phi_0(r)=-\frac{2}{3}\pi \rho_0 e r^2
\eeq
where $\phi_0(r)$ follows from Eq. (23b). The total energy of the electron is sum of
kinetic and potential energy:
\beq
E_{tot}=\frac{1}{2}m_e v^2-\frac{2}{3}\pi\rho_0 e r^2
\eeq
and is a constant of motion. Indeed, Eq. (46) is the same as Eq. (44b), with the integration constant
$k=2E_{tot}/m_e$. The allowed values of $k$ are such that the maximum elongation $r_{max}$ for
which $v$ goes to zero obeys  $r_{max}\leq R$. When $k$ is such that $r_{max}$ approaches $R$, the motion is slightly
faster than in the harmonic oscillator potential near the region of maximum elongation (as given by Eq. (44a)), due to screening
by the negative charge near the surface.
For small $R/\lambda_L$ instead the velocity is very different than Eq. (44b) and is given by
\beq
v^2=k'+\frac{2\pi e \rho_0}{15m_e\lambda_L^2}r^2[R^2-\frac{r^2}{2}]   
\eeq

Note also that an electron that moves in a harmonic oscillator potential
does not give rise to uniform charge density, but rather the electron density
and hence negative charge is largest near the region of maximum classical
elongation. This is consistent with the assumption that the charge density
in the superconductor is positive in the interior and negative near the surface. 
Similarly, circular 'Bohr orbits' in a harmonic oscillator potential 'bunch up' for large radius
as discussed in Ref. \cite{atom} giving rise to larger negative charge density near the surface.

The assumption $\vec{\nabla}\times \vec{v}=0$ implies linear motion where the electrons oscillate through
the origin. Relaxing this condition many other trajectories become possible. Deep in the interior the motion
is simply described by independent harmonic motion in the three directions with frequency given by
Eq. (39), which in general describes elliptical orbits. The linear oscillator and circular motion described above
are two particular examples of these orbits where it is simple to obtain the corrections due to the
outer layer of negative charge, as given by Eqs. (44a) and (37).

\section{Superconductors of general shape}

From Eq. (22a) we have
\beq
\tilde{\rho}(\vec{r})=\lambda_L\nabla^2 \tilde{\rho}(\vec{r})
\eeq
for $\tilde{\rho}(\vec{r})=\rho(\vec{r})-\rho_0$. Laue has shown rigurously\cite{laue} that for a body of 
arbitrary shape the solution of Eq. (48) decays to zero rapidly towards the interior, on the length scale
of the penetration depth $\lambda_L$. Hence we can conclude that quite generally for superconductors
of dimensions much larger than $\lambda_L$ the charge distribution in the interior is constant
and equal to $\rho_0$, and that the expelled negative charge resides within $\lambda_L$ of the
surface.

Similarly using that 
\beq
\vec{\nabla}\times(\vec{\nabla}\times\vec{E}_0)=\vec{\nabla}(\vec{\nabla}\cdot\vec{E}_0)-\nabla^2 \vec{E}_0
\eeq
as well as $\vec{\nabla}\times\vec{E}_0=0$ and $\vec{\nabla}\cdot\vec{E}_0=constant$ we conclude that
$\nabla^2 \vec{E}_0=0$, hence from Eq. (22b)
\beq
\nabla^2(\vec{E}-\vec{E}_0)=\frac{1}{\lambda_L^2}(\vec{E}-\vec{E}_0)
\eeq
This equation is the same as the London equation satisfied by the magnetic field in non-rotating
superconductors, giving rise to the Meissner effect. Again, Laue showed\cite{laue} that the
vector quantity satisfing Eq. (50) decays rapidly to zero over a distance of a penetration depth from
the surface. Hence we conclude that for superconductors
of arbitrary shape away form the surface the electric field is given by $\vec{E}_0$ originating in
the uniform charge distribution $\rho_0$. For example, for an ellipsoid of revolution with
symmetry axis along the $z$ direction
\beq
\frac{x^2+y^2}{a^2}+\frac{z^2}{b^2}=1
\eeq
the electric field is given by\cite{wangler}
\bmath
\beq
(E_0)_x=\frac{2}{3}\pi\rho_0(3-p)\frac{x}{a^2b}
\eeq
\beq
(E_0)_z=\frac{4}{3}\pi\rho_0p\frac{x}{a^2b}
\eeq
\emath
with $p=b/a$. Eq. (52) is approximately valid for $0.8<p<5$. In particular, the surface of a uniformly charged
ellipsoid is $not$ an equipotential surface, and this is of course also the case for a body of arbitrary
non-spherical shape.

In the presence of the expelled negative charge we argue that in general the surface of the body will still
not be an equipotential surface, and hence that electric fields will exist outside the superconductor. This is
seen as follows: we define
\beq
\tilde{\phi}(\vec{r})=\phi(\vec{r})-\phi_0(\vec{r})+4\pi \lambda_L^2\rho_0
\eeq
which obeys the differential equation
\bmath
\beq
\tilde{\phi}(\vec{r})=\lambda_L^2\nabla^2 \tilde{\phi}(\vec{r})
\eeq
inside the body, and
\beq
\nabla^2 \tilde{\phi}(\vec{r})=0
\eeq
\emath
outside. Furthermore, both $\tilde{\phi}$ and its normal derivative $\partial \tilde{\phi}/\partial n$ are
continuous on the surface of the body. The function $\tilde{\phi}$ satisfying the differential equation
(54) is uniquely determined by specifying $either$ $\tilde{\phi}$ $or$ $\partial \tilde{\phi}/\partial n$
on the surface of the body. For Eq. (54b) this is well known\cite{jackson},
for Eq. (54a) it follows similarly from Green's theorem since for any two solutions of 
Eq. (54a) with the same boundary values their difference $U(\vec{r})$ satisfies
\beq
\int_V d^3r [|\vec{\nabla}U|^2 +\frac{U^2}{\lambda_L^2}]=\oint_S U\frac{\partial U}{\partial n} da
\eeq
hence $U$ is identically zero in the interior. If no electric field existed outside we would have
 $\phi=constant$ and
$\partial \phi /\partial n=0$ on the surface of the body, and the function
$\tilde{\phi}$ would be overdetermined, i.e. there is in general no solution to the differential
equation (54) with both $\tilde{\phi}$ and $\partial \tilde{\phi} /\partial n$ specified on a 
closed surface.

To solve the problem in general numerically we suggest the following procedure. Assume an initial 
guess for $\phi(\vec{r})$ on the surface (for example constant). Solve the Dirichlet problem
for $\tilde{\phi}$ inside and outside, and find $\partial \tilde{\phi}/\partial n$ at the surface coming from the inside and
the outside, which in general will not match. Hence $\partial \phi/\partial n$ is discontinuous
at the surface, which implies the presence of a surface charge density
\beq
\sigma=\frac{1}{4\pi} [\frac{\partial \phi }{\partial n})_{inside} -\frac{\partial \phi }{\partial n})_{outside}]
\eeq
Since we assume no such surface charge is possible in superconductors it indicates that the initial guess for 
$\phi(\vec{r})$ on the surface was incorrect. We next compute the average of
$\partial \phi / \partial n$ inside and outside the surface, and solve the corresponding
Neumann problem for that boundary condition 
inside and outside. The resulting $\phi(\vec{r})$ on the surface will in general 
be discontinuous, so we calculate its average coming from inside and outside and solve 
again the Dirichlet problem with the new boundary condition. Thus solving a sequence of alternating
Dirichlet and Neumann problems the solution should converge to a unique solution with both
$\phi$ and $\partial \phi / \partial n$ continuous at the surface.

These considerations also indicate that for superconductors of dimensions much larger than the
penetration depth the negative charge near the surface  will  arrange laterally so as
to nullify any tangential electric field and hence that the surface of the body will be very nearly
an equipotential surface. This is because the surface charge density Eq. (56) required for it
can be achieved by arranging the charge in the surface $layer$ of thickness $\lambda_L$ 
appropriately. In other words, the solution to the differential equations where the surface charge
$\sigma$ in Eq. (56) is spread out over the layer of thickness $\lambda_L$ and
$\phi$ is very nearly constant on the surface will be the unique solution for a body of dimensions
much larger than $\lambda_L$. If $\phi$ is constant on the surface and the body is charge neutral,
no electric field lines can exist outside: for any exterior electric field line starting on the surface would
have to return to another point on the surface, and the integral of the electric field along that line
would yield a difference in potential at the initial and final points, both on the surface.
Hence we conclude
that for a charge neutral superconductor of large dimensions essentially no electric
field lines will exist outside the superconductor, and it is not possible to detect the
non-uniform charge distribution in the interior by measurements outside the superconductor.

Instead, for superconductors of dimensions comparable to or smaller than the
penetration depth the negative charge distribution cannot be regarded as a
'surface charge', and it   will not  cancel the
potential differences at different points of the surface originating in the positive charge.
Hence the electric field will 'leak out' to the exterior of the superconductor and
become observable.

The simplest shape that can give rise to such effect is an ellipsoid of revolution, described by Eq. (51). For a uniformly
charged ellipsoid with charge density $\rho_0$ the dipole moment (with respect to the center of mass) is zero, but
the quadrupole moment is given by
\beq
Q\equiv \int d^3r \rho(\vec{r}) (3z^2-r^2)=\frac{8}{15}\pi a^2b(b^2-a^2)\rho_0
\eeq
For the charge neutral superconductor, if negative charge is expelled towards the surface the charge
distribution looks qualitatively as in Figure 3. If the dimensions of the body are not much larger than
the penetration depth it is plausible to assume that 
the negative charge will not rearrange laterally  
 but instead will remain approximately uniform. It is clear from Fig. 3 that the negative
charge will give a larger contribution to the quadrupole moment than the positive charge. 
Consequently this will give rise to a net $negative$ quadrupole moment
for a prolate ellipsoid ($b>a$ with $b$ in the $z$ direction), and
a net $positive$ quadrupole moment  for
an oblate ellipsoid ($b<a$).

\begin{figure}
  \includegraphics[height=.22\textheight]{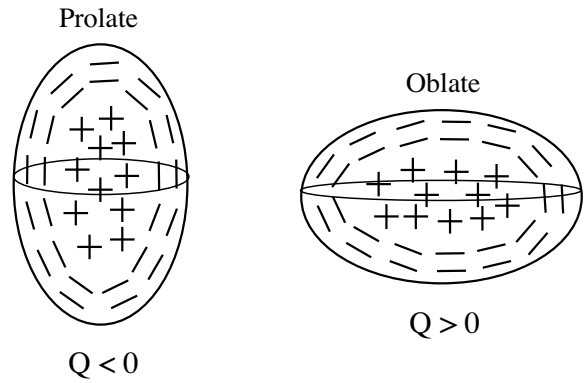}
  \caption{Charge density in ellipsoids of revolution of dimensions comparable to the penetration depth (schematic). 
There will be a net quadrupole moment (negative for prolate, positive for oblate) giving rise to an electric field
outside the superconductor whose magnitude decays as the fourth power of the distance to the center of the ellipsoid.
}
\end{figure}

For body dimensions comparable but somewhat larger than the penetration depth we can estimate the
quadrupole moment assuming charge density $\rho_-$ within distance
$\lambda_L$ of the surface and $\rho_0$ in the interior. The net quadrupole
moment that results is
\beq
Q=\frac{16}{15}\pi a^2b(b-a)\lambda_L\rho_-
\eeq
so that it is negative (positive) for a prolate (oblate) ellipsoid.
The electric field at distance $r$ from the center $outside$ the body  is of order
$E\sim Q/r^4$ and should be measurable.

For a body of dimensions much smaller than the London
penetration depth we can   estimate the net quadrupole moment  as follows. The charge density Eq. (30) becomes for the ellipsoid
\beq
\rho(r)=\frac{\rho_0}{6\lambda_L^2} [\frac{2a^2+b^2}{5}-r^2]
\eeq
to ensure charge neutrality, 
and calculating the integral Eq. (57) over the volume of the ellipsoid yields
\beq
Q=-\frac{8}{1575} \pi a^2b (b-a)(3a^3+3a^2b+4ab^2+4b^3)\frac{\rho_0}{\lambda_L^2}
\eeq
For small distortion from sphericity in terms of $R\sim a \sim b$
\beq
Q=-\frac{16}{15}\pi a^2b(b-a)\lambda_L\times[\frac{2}{15}\frac{\rho_0R^3}{\lambda_L^3}]
\eeq
 which is similar  to Eq. (58) for $R\sim \lambda_L$.
If we relate $\rho_0$ to $\rho_-$ using Eq. (59) with $\rho_-\equiv \rho(r=R)$ Eqs. (61) and (58) differ 
by a factor $4/3$.

\begin{figure}
 \includegraphics[height=.25\textheight]{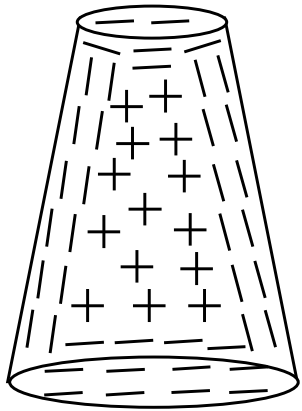}
  \caption{Charge density in a body of low enough symmetry that it can give rise to an electric dipole moment  (schematic). 
For dimensions of the body comparable to or smaller than the penetration depth there will be 
a net electric dipole moment along the vertical direction giving rise to a dipolar electric field outside
the superconductor.}
\end{figure}

For body shapes with less symmetry than the ellipsoid, a dipole moment also becomes  possible.
As an example, consider the body of revolution shown in Figure 4, with radius  
\beq
r(z)=a+(b-a)\frac{z}{h}
\eeq
and $0\leq z\leq h$. For a uniformly charged such body with  charge density  $\rho$ the dipole moment
relative to $(r=0, z=h/2)$ is given by
\beq
p=\frac{\pi h^2}{12}(b^2-a^2)\rho
\eeq
We assume again that the expelled negative charge does not rearrange laterally for body
dimensions not much larger than the penetration depth. The net dipole moment is then given by
\beq
p =\frac{\pi}{12}\rho_- h[h(b^2-a^2)-h'(b'^2-a'^2)\frac{a^2+ab+b^2}{a'^2+a'b'+b'^2}]
\eeq
with $h'=h-2\lambda_L$, $a'=r(\lambda_L)-\lambda_L$, $b'=r(h-\lambda_L)-\lambda_L$, 
and points along the $z$ direction. The sign
of $p$ is not simply determined, e.g. for $b<a$ it is generally  positive   but can become negative for
large $h$.

\section{Discussion}
We have proposed that the ground state charge distribution in superconductors is
 different from what is conventionally assumed. Rather than being
uniform and locally charge neutral as normal metals, we propose that the charge distribution in
superconductors is inhomogeneous, with more negative charge near the surface
and more positive charge in the interior, reflecting the
fundamental charge asymmetry of matter.

The charge inhomogeneity is a consequence of the fact that the superfluid electrons are 
highly mobile and have quantum-mechanical coherence over the macroscopic
dimensions of the sample. These electrons have 'undressed' from the
electron-ion and the electron-electron interaction\cite{undressing, atom}
 and become completely 
free-electron-like, except for the pairing correlations that bind $k\uparrow$ and
$-k\downarrow$ electrons. The quantum-mechanical lowering of kinetic energy of
the light electrons due to delocalization causes the electronic charge density to be
larger near the boundaries of the sample, just as electrons do not remain confined
within the dimensions of the positive nucleus in an ordinary atom. 

We argue that the electric field in the interior of superconductors is not screened as in
ordinary metals because the ground state wave function has 'rigidity', and the
macroscopic quantum-mechanical coherence prevents local deformations that would 
screen the electric field. As discussed in Sect. 4, an electric field can exist in a 
superconductor if the superfluid electrons are delocalized over macroscopic
distances and their velocity has the proper gradient.
 At finite temperatures however there are also excited
quasiparticles that are not  macroscopically coherent, and one may wonder 
why they do not screen the electric field. We suggest that they are
unable to screen the interior field because they are positively charged\cite{thermo} and
as a consequence they are also pushed out towards the surface.

In the Meissner effect, the supercurrents can shield the magnetic field so that it is zero in the interior of the superconductor
only if the body's dimensions are larger than the penetration depth; otherwise, the magnetic field penetrates the 
superconductor. Similarly, we have argued that
the negative charge can shield the electric field so that it is 
zero in the $exterior$ of the superconductor if the body's dimensions are larger than the penetration depth; otherwise, the electric
 field can $leak$ $out$ from  the superconductor to the exterior, unless the
body has perfect spherical symmetry.

The latter fact should allow for experimental detection of these electric fields around small superconducting
particles of non-spherical shape.  Remarkably, 
very recently  Moro et al reported detection of  spontaneous electric dipole moments  in Nb clusters\cite{clusters} at
low temperatures. Their observations that the effect is strongest for $even$ number of electrons in the cluster and
that it occurs only at low temperatures suggests that the effect is related to superconductivity\cite{clusters} as proposed by the authors. 
Moro et al concluded from their observations the existence of an internal
electric field in the interior of the metal cluster of the order of $10^6 V/cm$, and stress  that such an
internal electric field cannot occur in an ordinary metal and suggests a 'rigidity' of the electronic
wavefunctions as well as a collective effect. 
The theory discussed here predicts that  dipole moments should arise  from clusters of irregular shape such as
shown in Fig. 4. Experimentally, relation between the measured dipole moment and the shape of the
cluster has not been  examined, and it would be interesting to do so to compare with
predictions of this theory. Concerning the magnitude of the expected effect, we estimated in
Sect. (1) a maximum electric field for $Nb$ comparable to the one inferred by Moro et al under
the assumption that the energy $\tilde{\epsilon}$ in Eq. (7) is the condensation energy. 
However the magnitude of the electric field will be much smaller for cluster dimensions smaller
than the penetration depth according to the results in Sect. III, so that our prediction may be
inconsistent with the observation unless $\tilde{\epsilon}$ is assumed to be much larger than
the condensation energy. 

For small clusters of more regular shape such as ellipsoids of revolution, no dipole moment  
but a quadrupole moment should be observed, as discussed earlier. Experimental determination of
quadrupole moments in superconducting metal clusters has not  been reported, and it would be 
interesting to search for this effect to compare with the theory discussed here. For example, it has been
reported that tin clusters adopt prolate geometries\cite{tin}. If so, according to the theory discussed
here  tin clusters at low temperatures should exhibit a $negative$ quadrupole moment.

Further consequences of this physics will be discussed in  future work.

\acknowledgements
The author is grateful to Michel Viret for calling ref. \cite{clusters} to his attention, and to 
W.A. de Heer for a stimulating discussion.


\begin{references}
\bibitem{explain} In ref. \cite{atom} it was also found that some of the negative charge 
spills out beyond the surface of the superconductor as indicated in Fig. 1; 
however that effect will not be considered here.
\end{references}
\end{document}